\documentclass[fleqn,twoside]{article}
\usepackage{espcrc2}

\usepackage{graphicx}

\newcommand {\nn}{\nonumber \\}

\newcommand {\e}{{\rm e}}

\newcommand {\m}{\mu}
\newcommand {\n}{\nu}
\newcommand {\pl}{\partial}

\newcommand {\al}{\alpha}
\newcommand {\be}{\beta}
\newcommand {\ga}{\gamma}

\newcommand {\la}{\lambda}
\newcommand {\La}{\Lambda}
\newcommand {\si}{\sigma}
\newcommand {\Si}{\Sigma}

\newcommand {\om}{\omega}

\newcommand {\ep}{\epsilon}

\newcommand {\del}  {\delta}

\newcommand {\mn}{{\mu\nu}}
\newcommand {\ls}   {{\lambda\sigma}}
\newcommand {\ab}   {{\alpha\beta}}
\newcommand {\gd}   {{\gamma\delta}}
\newcommand {\half}{ {\frac{1}{2}} }

\newcommand {\fourth} {\frac{1}{4} }

\newcommand {\Lcal}{{\cal L}}

\newcommand {\Rhat}{{\hat R}}

\newcommand {\ehat}{{\hat e}}
\newcommand {\mhat}{{\hat m}}

\newcommand {\etahat} {{\hat \eta}}
\newcommand {\omhat} {{\hat \omega}}
\newcommand {\psihat} {{\hat \psi}}
\newcommand {\thhat} {{\hat \theta}}
\newcommand {\gh}  {{\hat g}}

\newcommand {\psibar}{{\bar \psi}}

\newcommand  {\vz}{{v_0}}

\newcommand {\ra} {\rightarrow}

\newcommand {\pr}   {{\quad .}}
\newcommand {\com}  {{\quad ,}}
\newcommand {\q}    {\quad}

\newcommand {\NP}   {Nucl.Phys.}

\newcommand {\PRL}   {Phys.Rev.Lett.}

\newcommand {\gago}  {\gamma_5}

\newcommand{\AmS}{{\protect\the\textfont2
  A\kern-.1667em\lower.5ex\hbox{M}\kern-.125emS}}

\title{
CP-Violation in Kaluza-Klein and Randall-Sundrum Theories
}

\author{S. Ichinose
   \address
{Laboratory of Physics, 
School of Food and Nutritional Sciences, \\
University of Shizuoka,
Yada 52-1, Shizuoka 422-8526, Japan 
}%
        \thanks{E-mail address:\ ichinose@u-shizuoka-ken.ac.jp}
        }
       
\begin{document}

\begin{abstract}
The Kaluza-Klein theory and Randall-Sundrum
theory are examined comparatively, with focus
on the five dimensional (Dirac) fermion
and the dimensional reduction to four dimensions.
They are treated in the Cartan formalism. 
The chiral property, localization, anomaly phenomena
are examined. The electric and magnetic dipole
moment terms naturally appear. The order estimation
of the couplings is done. This is a possible
origin of the CP-violation.
\vspace{1pc}
\end{abstract}

\maketitle

\section{Introduction and Conclusion}
If our present research direction 
of the string and D-brane is right, 
the unification of various forces should be, 
effectively at some scale, explained by some higher-dimensional
theory. 
Then it is quite sure that
the real world of 4 dimensions should be some approximation
of the higher dimensional one. It is realized by 
the procedure called {\it dimensional 
reduction}. There are two 
representative and contrastive
approaches, that is, the Kaluza-Klein and the Randall-Sundrum
theories. 
In the former case, the reduction is achieved by
the {\it compactification} of the extra space, while
in the latter one it is done by the {\it localization}
of the configuration along the extra dimension(s). 
The two approaches look to have
both advantages and disadvantages in the phenomenological 
application. Here we treat them in a comparative way and examine
their 4 dimensional(D) reduction properties. 

We will present the higher dimensional approach
to the CP-violation mechanism. As was stressed by
Thirring for the KK model\cite{Thirr72}, the CP-violation naturally
occurs also in the RS model.


\section{Fermions in Kaluza-Klein Theory}
Let us first review the 5D Kaluza-Klein theory.
This serves as the preparation for the same treatment of 
the Randall-Sundrum theory in the next section.
The 5D space-time manifold is described by the 4D coordinates
$x^a$ ($a=0,1,2,3$) and an extra coordinate $y$. We also use
the notation ($X^m$)=($x^a, y$), ($m=0,1,2,3,5$). 
With the general 5D metric $\gh_{mn}$:\ 
$ ds^2=\gh_{mn}(X)dX^mdX^n $
, we assume the $S^1$ compactification condition for the
extra space:\ 
$ \gh_{mn}(x,y)=\gh_{mn}(x,y+{2\pi}/{\mu}) $
, where $\mu^{-1}$ is the {\it radius} of the extra space circle.
We specify the form of the metric as
\begin{eqnarray}
ds^2
=g_{ab}(x)dx^adx^b+\e^{\si(x)}(dy-fA_a(x)dx^a)^2
\com
\label{KK1c}
\end{eqnarray}
where $g_{ab}(x), A_a(x)$ and $\si(x)$ are 
the 4D metric, the U(1) gauge field and the dilaton field
respectively. 
$f$ is a coupling constant. 
This specification is based on the
following additional assumptions:\ 
1.
$y$ is a space coordinate;\ 
2.
The geometry is invariant under
the U(1) symmetry, 
$y\ra y+\La(x),\ A_a(x)\ra A_a(x)+\frac{1}{f}\pl_a\La$.
We take $\si(x)=0$ in (\ref{KK1c}) for simplicity.

We take the Cartan formalism to
compute the geometric quantities\cite{Thirr72}. 
The basis \{$\thhat^\m$\} 
($\m= 0,1,2,3,5$ are the local Lorentz (tangent) frame indices)
of the cotangent manifold($T_p^*M$)
and the f\"{u}nf-bein $\ehat^\m_{~m}$
are obtained as
\begin{eqnarray}
\thhat^\al=\theta^\al=e^\al_{~a}dx^a,\ \thhat^5=dy-fA_adx^a,
A_\al\equiv e_\al^{~a}A_a,\nn
(\ehat^\m_{~m})=\left(
\begin{array}{cc}
e^\al_{~a} & 0 \\
-fA_a   & 1
\end{array}
                 \right),\ 
(\ehat_\m^{~m})=\left(
\begin{array}{cc}
e_\al^{~a} & 0 \\
fA_\al   & 1
\end{array}
                 \right),
\label{KK5}
\end{eqnarray}
where $\theta^\al$ and $e^\al_{~a}$ ($\al=0,1,2,3$)
are the 4D part of $\ehat^\m_{~m}$ and $\thhat^\m$
respectively. 
The first Cartan's structure equation gives
the connection 1-form $\omhat^\m_{~\n}$,
\begin{eqnarray}
\omhat^5_{~5}=0,\ \omhat^5_{~\al}=-\frac{f}{2}F_\ab\theta^\be,
\nn
\omhat^\al_{~\be}=\om^\al_{~\be}
-\frac{f}{2}F^\al_{~\be}\thhat^5,\ 
F_\ab\equiv e_\al^{~a}e_\be^{~b}F_{ab}
\com
\label{KK9}
\end{eqnarray}
where $\om^\al_{~\be}$ is the 4D part.

The 5D Dirac equation is generally given by
\begin{eqnarray}
\left\{
\ga^\m\ehat_\m^{~m}\frac{\pl}{\pl X^m}
+\frac{1}{8}(\omhat^\si)_\mn\ga_\si [\ga^\m,\ga^\n]-\mhat
\right\}\psihat=0
.
\label{ferKK1}
\end{eqnarray}
The spin connection above $(\omhat^\si)_\mn$ is defined by
$\omhat^\mu_{~\nu}=(\omhat_\la)^\mu_{~\nu}\thhat^\la$.
The 5D Dirac matrix $\ga^\m$ satisfies
$\{\ga^\m,\ga^\n\}=2\etahat^\mn,
\etahat^\mn=\mbox{diag}(-1,1,1,1,1)$.
For simplicity we switch off the 4D gravity: 
$e^\al_{~a}\ra\del^\al_a\ ,\ \om^\al_{~\be}\ra 0$. 
The parameter $\mhat$ is the 5D fermion mass.
In the present case, using (\ref{KK5}) and (\ref{KK9}),
the eq.(\ref{ferKK1}) says
\begin{eqnarray}
\left\{
\ga^a(\pl_a+fA_a\pl_5)+\ga^5\pl_5\right.\nn
\left.
+\frac{f}{16}F_{ab}\ga^5[\ga^a,\ga^b]-\mhat
\right\}\psihat=0
\pr
\label{ferKK2}
\end{eqnarray}

We consider the following form of the fermion:\ 
$ \psihat(x,y)=\e^{i(\phi\gago+\m y)}\psi(x) $
. Here we regard the fermion as a KK-{\it massive} mode.
The angle parameter $\phi$ is chosen as
\begin{eqnarray}
(i\gago\m+\mhat)\e^{2i\phi\gago}
=\sqrt{{\mhat}^2+\m^2}\equiv M,\ 
\tan 2\phi=-\frac{\m}{\mhat}
.
\label{ferKK4}
\end{eqnarray}
Then (\ref{ferKK2}) reduces to
\begin{eqnarray}
\left\{
\ga^a(\pl_a+ieA_a)-M
+\right.\nn
\left.
\frac{1}{16M}(\mhat\frac{e}{\m}-ie\gago)F_{ab}\ga^5[\ga^a,\ga^b]
\right\}\psi=0
\com
\label{ferKK5}
\end{eqnarray}
where $e\equiv f\m$ is the electric coupling constant.
We notice, in this expression, 
the electric and  magnetic moments appear.
\begin{eqnarray}
\frac{1}{16}\frac{\mhat}{M\m}eF_{ab}\psibar\ga^5[\ga^a,\ga^b]\psi,\ 
-\frac{i}{16}\frac{1}{M}eF_{ab}\psibar[\ga^a,\ga^b]\psi
.
\label{ferKK6}
\end{eqnarray}
The first term violate the CP symmetry. 
Note that the second term usually appear, in the 4D QED, 
as the {\it quantum} effects. 

Let us do the order estimation. From the reduction
of $
\int\sqrt{-\gh}\Lcal^{grav}=\frac{1}{G_5}\int d^5X\sqrt{-\gh}\Rhat
$,
\begin{eqnarray}
\frac{1}{G_5}\frac{2\pi}{\m}\int d^4x 
\sqrt{-g}(R-\frac{f^2}{4}F^\ab F_\ab+\cdots)
,
\label{ferKK7}
\end{eqnarray}
we know
$ {G_5\m}\sim {G},\ 
{f^2}/{G_5\m}\sim 1 $
, where $G$ is the gravitational constant. 
This gives $f\sim \sqrt{G}=10^{-19}\mbox{GeV}^{-1}$.
On the other hand, we know $e=\m f\sim 10^{-1}$.
Hence we obtain $\m\sim 10^{-1}f^{-1}\sim 10^{18}\mbox{GeV}$.
Assuming $\mhat\sim\m$, we can estimate 
the electric and magnetic couplings as
\begin{eqnarray}
\frac{\mhat}{M\m}e\sim
\frac{e}{M}\sim 10^{-19}\mbox{GeV}^{-1}\sim 10^{-32}\mbox{e cm}
\com
\label{ferKK9}
\end{eqnarray}
which is {\it far below} the experimental upper bound
of the neutron electric dipole moment $6.3\times 10^{-26}$e cm.

\section{Fermions in Randall-Sundrum Geometry}
We consider the following 5D space-time geometry\cite{RS9905}.
\begin{eqnarray}
ds^2=\e^{-2\si(y)}\eta_{ab}dx^adx^b+dy^2
=\gh_{mn}dX^mdX^n,\nn
-\infty<y<+\infty\ ,\ -\infty<x^a<+\infty
,
\label{RS1}
\end{eqnarray}
where $\si(y)$ is regarded as a scale factor field.
$(\eta_{ab})=\mbox{diag}(-1,1,1,1)$.
When the geometry is AdS$_5$, $\si(y)=c|y|, c>0$. 
Such a situation, in the present case, 
occurs in the asymptotic region
of the extra space $y\ra\pm\infty$. 
The 1-form $\thhat^\m$, 
the f\"{u}nf-bein $\ehat^\m_{~m}$ and
the connection 1-form $\omhat^\m_{~\n}$
are given by
\begin{eqnarray}
\thhat^\al=\e^{-\si}\eta^\al_{~a}dx^a\com\q
\thhat^5=dy
\com\nn
(\ehat^\m_{~m})=\left(
\begin{array}{cc}
e^{-\si}\eta^\al_{~a} & 0 \\
0   & 1
\end{array}
                 \right),\ 
(\ehat_\m^{~m})=\left(
\begin{array}{cc}
e^\si\eta_\al^{~a} & 0 \\
0   & 1
\end{array}
                 \right)
                 ,\nn
\omhat^5_{~5}=0\com\q 
\omhat^\al_{~5}=-\omhat_5^{~\al}=-\si'\theta^\al\com\nn
\omhat^5_{~\al}=-\omhat_\al^{~5}=\si'\theta_\al\com\q
\omhat^\al_{~\be}=0\com
\label{RS3}
\end{eqnarray}
where $\si'=\frac{d\si}{dy}$.

As the fermion model, 
the (5D) Dirac theory (\ref{ferKK1}), $\sqrt{-\gh}\Lcal^{Dirac}$, 
is not accepted physically 
because the fermion configuration does not localize (around
$y=0$) in the extra space.
Nature requires the {\it Yukawa interaction} 
between the 5D fermion and the 5D Higgs\cite{BG99}.

Hence we examine the fermion behavior
under the Yukawa coupling with the Higgs scalar:\ 
$
\sqrt{-\gh}\Lcal=\sqrt{-\gh}(\Lcal^{Dirac}+\Lcal^Y),\ 
\Lcal^Y=ig_Y{\bar \psihat}\psihat\Phi
,
$
where $\Phi$ is the 5D(bulk) Higgs scalar field
and $g_Y$ is the Yukawa coupling. This reduces to
\begin{eqnarray}
i\e^{\si}\{
\ga^a\pl_a+2\e^{-\si}(\si'-\half\pl_y)\gago-g_Y\e^{-\si}\Phi
                           \}\psihat=0
.
\label{yuka5}
\end{eqnarray}
A solution of the right chirality zero mode:\ 
$
\psihat(x,y)=\psi^0_R(x)\eta(y),\ 
\gago\psi^0_R=+\psi^0_R,\ 
\ga^a\pl_a\psi^0_R=0
$
, is obtained by
\begin{eqnarray}
\pl_y\eta=(2\si'-g_Y\Phi(y))\eta
\pr
\label{yuka7}
\end{eqnarray}
In the far region, the solution behaves as
$ \eta(y)=\mbox{const}\times \e^{-(g_Y\vz-2\om)|y|} $
, which shows the {\it exponentially damping} for the large
yukawa coupling $g_Y>2\om/\vz$. 
($\om$ and $\vz$ is the asymptotic values of $\si'(y)$
and $\Phi(y)$\cite{SI0003}.
)
This is called
{\it massless chiral fermion localization}. In the near region, 
$ \eta(y)=\mbox{const}\times \e^{-\frac{k}{2}(g_Y\vz-2\om)y^2} $
, which shows the {\it Gaussian damping}. 
($1/k$ is the {\it thickness} parameter of the wall.
)

Now we analyze the 5D QED, 
$\Lcal^{QED}=-e{\bar \psihat}
\ga^\m\ehat_\m^{~m}\psihat A_m$, 
with the Yukawa interaction in RS geometry:\ 
$ \sqrt{-\gh}(\Lcal^{Dirac}+\Lcal^{EM}+\Lcal^{QED}+\Lcal^Y), 
\Lcal^{EM}=-\fourth \gh^{mn}\gh^{kl}F_{mk}F_{nl} $
.
We assume, as in the previous paragraph, $\gh^{mn}$ and $\Phi$ are
the brane background obtained from the gravitational and Higgs systems:
$\sqrt{-\gh}(\Lcal^{grav}+\Lcal^S)$.

Let us examine the {\it bulk quantum} effect. It induces the 5D 
effective action $S_{eff}$ which reduces to the 4D action
in the {\it thin wall limit}. 
From the diagram of Fig.1(left), we expect
\begin{eqnarray}
\frac{\del S^{(1)}_{eff}}{\del A^\mu(X)}\equiv
<J_\m>\sim e^2g_Y\ep_{\mn\ls\tau}\Phi F^{\nu\la}F^{\si\tau}
\pr
\label{qed3}
\end{eqnarray}
\begin{figure}[htb]
\includegraphics[height=5pc]{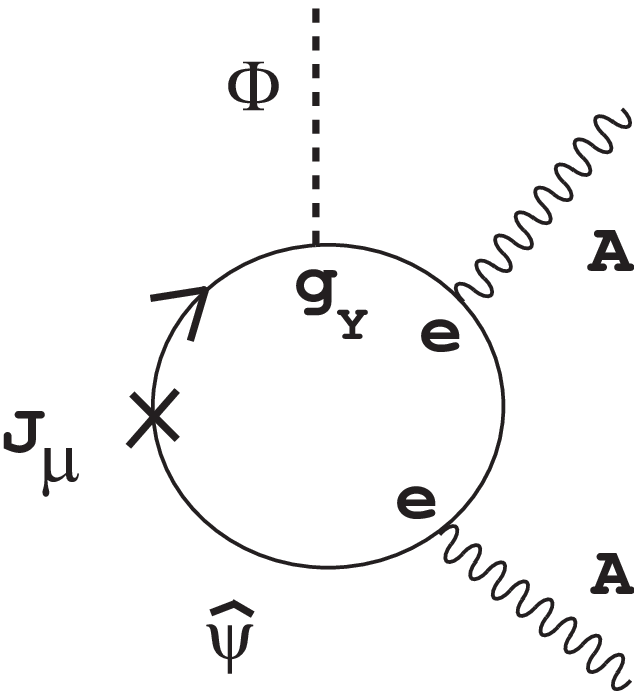}\hspace{30pt} 
\includegraphics[height=5pc]{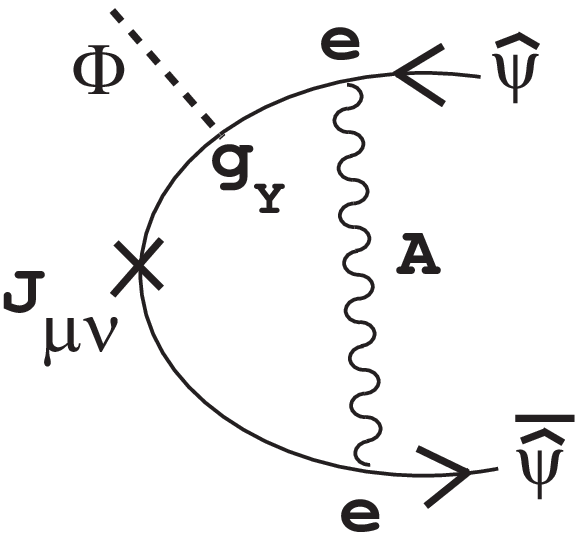}
\caption{The left diagram induces the effective action
$S^{(1)}_{eff}$, while the right one induces $S^{(2)}_{eff}$.}
\label{fig:anomaly}
\end{figure}
Then the effective action is integrated as
\begin{eqnarray}
S^{(1)}_{eff}\sim
e^2g_Y\int d^5X\ep_{\mn\ls\tau}\Phi A^\m F^{\nu\la}F^{\si\tau}
\pr
\label{qed4}
\end{eqnarray}
In the {\it thin wall limit} we may approximate as
$\Phi=\Phi(y)\sim \vz\ep(y)$ where $\ep(y)$ is the step function.
Under the U(1) gauge transformation $\del A^\m=\pl^\m\La$, 
$S^{(1)}_{eff}$ changes as
\begin{eqnarray}
\del_\La S^{(1)}_{eff}\sim
e^2g_Y\vz\int d^5X\ep_{\mn\ls\tau}\ep(y) \pl^\m\La
F^{\nu\la}F^{\si\tau}\nn
=-e^2g_Y\vz\int d^4x
\La(x) F^{\ab}{\tilde F}_{\ab}
\com
\label{qed5}
\end{eqnarray}
where ${\tilde F}_{\ab}\equiv \ep_{\ab\ga\del}F^{\ga\del}$. 
In the above we assume that the boundary term vanishes. 
Callan and Harvey interpreted this result as 
the "anomaly flow" between
the boundary (our 4D world) and the bulk\cite{CH85}. 

Through the analysis of 
the induced action in the bulk,
we can see the "dual" aspect of the 4D QED. 

Another interesting bulk quantum effect is given by Fig.1(right). 
The induced effective action $S^{(2)}_{eff}$ is expected
to satisfy
\begin{eqnarray}
\frac{\del S^{(2)}_{eff}}{\del F^\mn}\equiv
<J_\mn>\sim e^2g_Y\ep_{\mn\ls\tau}\pl^\la\Phi 
{\bar \psihat}\Si^{\si\tau}\psihat
\pr
\label{qed6}
\end{eqnarray}
Then $S^{(2)}_{eff}$ is obtained as, in the thin wall limit, 
\begin{eqnarray}
S^{(2)}_{eff}
\sim e^2g_Y\ep_{\mn\ls\tau}\int d^5X\pl^\la\Phi F^\mn
{\bar \psihat}\Si^{\si\tau}\psihat\nn
= e^2g_Y\vz\ep_{\ab\gd}\int d^4x F^\ab
{\bar \psi}\si^{\gd}\psi
\pr
\label{qed7}
\end{eqnarray}
This term is the "dual" of the magnetic moment term of
(\ref{ferKK6}). It is a CP-violating term. The coupling
depends on the vacuum expectation value of $\Phi$.


\end{document}